\documentclass[%
 reprint,
 amsmath,amssymb,
 aps,
 nofootinbib,
]{revtex4-2}

\def\babar{\mbox{\slshape B\kern-0.1em{\smaller A}\kern-0.1em
    B\kern-0.1em{\smaller A\kern-0.2em R}}}

\usepackage[export]{adjustbox}
\usepackage{relsize}
\usepackage{graphicx}
\usepackage{dcolumn}
\usepackage{bm}
\usepackage{graphicx}
\usepackage{comment}
\usepackage[caption=false]{subfig}
\usepackage{rotating}
\usepackage{lipsum}
\usepackage[]{hyperref}
\hypersetup{
colorlinks,
linkcolor={blue},
linktocpage=true,
citecolor={red},
urlcolor={green}
}

\begin{document}
\babar-PUB-23/03 \\
\vspace{0.3cm} 
 \hspace{0.15cm} SLAC-PUB-17731 \\

\title{Search for Evidence of Baryogenesis and Dark Matter in $B^{+} \rightarrow \psi_{\text{D}} + \text{p}  $ Decays at \babar}


\begin{abstract}
 A new dark sector anti-baryon, denoted $\psi_{\text{D}}$, could be produced in decays of $B$ mesons. This letter presents a search for $B^{+} \rightarrow \psi_{\text{D}} + \text{p}$ (and the charge conjugate) decays in $e^{+}e^{-}$ annihilations at 10.58 GeV, using data collected in the ~\babar~ experiment. Data corresponding to an integrated luminosity of 398 fb$^{-1}$ are analyzed.   No evidence for a signal is observed. Branching fraction upper limits in the range 
    from $10^{-7} $ -- $ 10^{-5}$ are obtained at 90$\%$ confidence level for masses of 1.0 $< m_{\psi_{\text{D}}} < $ 4.3 GeV/c$^{2}$. The result is also reinterpreted to provide the first limits on a supersymmetric model with R-parity violation and a light neutralino.

\end{abstract}

\author{J.~P.~Lees}
\author{V.~Poireau}
\author{V.~Tisserand}
\author{E.~Grauges}
\author{A.~Palano}
\author{G.~Eigen}
\author{D.~N.~Brown}
\author{Yu.~G.~Kolomensky}
\author{M.~Fritsch}
\author{H.~Koch}
% ...
\collaboration{The \babar\ Collaboration}
\noaffiliation

\maketitle


The existence of dark matter (DM) is established from astrophysical observations \cite{lensing, Rubin:1980zd,CMB}. Measurements of the cosmic microwave background (CMB) by the Planck satellite \cite{refId0} have shown that only $\sim$15$\%$ of the matter content of the universe can be accounted for from Standard Model (SM) particles. The remaining fraction is referred to as DM. Understanding the mass scale and nature of DM is one of the most pressing issues of modern particle physics. 

Another pressing issue is understanding the baryon asymmetry of the universe (BAU) \cite{Canetti_2012}. A dynamical mechanism, baryogenesis, is required to produce an initial excess of baryons over anti-baryons consistent with CMB and big-bang nucleosynthesis (BBN) measurements \cite{PDG2020,RevModPhys.88.015004}. 

 In Ref. \cite{Elor:2018twp} a new dark sector anti-baryon\footnote{The charge conjugate involves a dark sector baryon accompanied by an anti-proton, both channels are used in this analysis. This is implied throughout.}, $\psi_{\text{D}}$, is proposed, which can also explain the BAU.  In this model, baryogenesis occurs due to out-of-thermal-equilibrium production of $b$ and $\bar{b}$ quarks in the early universe through the decay of a massive, long-lived scalar field. The $b$ and $\bar{b}$ quarks hadronize into  $B^{0}_{s}, B^{0}$, and $ B^{\pm}$ mesons. The $B^{0} $ -- $ \bar{B}^{0}$ mesons then undergo CP-violating oscillations before decaying into a SM baryon $\mathcal{B}$, $\psi_{\text{D}}$, and any number of additional light mesons $\mathcal{M}$.  These CP-violating oscillations can originate from the SM or beyond the standard model (BSM) processes.  The term $B$-mesogenesis is coined to describe this mechanism. Decays of $B$ mesons into $\psi_{\text{D}}$ are mediated by new particles introduced at the TeV scale. In this scenario, matter-antimatter asymmetries are generated in the visible and dark sectors with equal magnitudes but opposite signs, keeping the total baryon number conserved. Current bounds on the semi-leptonic charge asymmetry in the decays of $B^{0}_{s}$ and $ B^{0}$ set a lower bound  on the total branching fraction $BF(B \rightarrow {\cal B}\psi_{\text{D}} {\cal M})  \gtrsim  10^{-4}$ assuming 
that the observed baryon-antibaryon-asymmetry is explained solely by the
mesogenesis mechanism of Ref.~\cite{Alonso-Alvarez:2021qfd}.

We present herein a search for the exclusive decay $B^{+} \rightarrow \psi_{\text{D}} + \text{p}$ and its charge conjugate. We utilize the hadronic recoil $B$-tagging method as outlined in Ref.~\cite{BaBar:2014omp}. One of the $B$ mesons from $e^{+}e^{-}\rightarrow B^{+}B^{-}$ is fully reconstructed from known hadronic decay modes, and is referred to as the $B_{\text{tag}}$ \cite{BaBar:2014omp}. The rest of the event\footnote{All other tracks and clusters in the event.}, which must include the proton, is then assigned to the other $B$ meson, denoted as the $B_{\text{sig}}$. Previous limits have been provided from a reinterpretation of a search for decays of $b$-flavored hadrons with large missing energy at LEP \cite{Alonso-Alvarez:2021qfd,ALEPH:2000vvi}. In addition, direct searches for the TeV-scale mediator at the LHC~\cite{CMS2019,ATLAS2020}, and DM stability, require $0.94 < m_{\psi_{\text{D}}} < 3.5 $ GeV/c$^{2}$~\cite{Alonso-Alvarez:2021qfd}. 

Constraints on exclusive decays (with a single SM baryon in the final state) are calculated using phase-space considerations for different baryons \cite{Alonso-Alvarez:2021qfd}. The results depend on the effective operators $\mathcal{O}_{i,j}=(\psi_{\text{D}}b) (q_{i}q_{j})$ mediating the decay, where $i$ and $j$ specify the quark content, $q_{i} = u, c$ and $q_{j} = d, s$. There are four possible flavor-combination operators of interest for $B$ meson decays.  The decay presented here probes $\mathcal{O}_{ud}$. New limits on $B^{0} \rightarrow \psi_{\text{D}} + \Lambda$ from ~\babar~ are presented in Ref. \cite{BaBar:2023rer}, which probes the $\mathcal{O}_{us}$  operator. Since the presented search is not sensitive to the Dirac or Majorana nature of the \text{invisible} particle, it is potentially sensitive to other models predicting $B^{+} \rightarrow \text{invisible} + \text{p}$ . In our conclusion, we also reinterpret the search for a supersymmetric model with R-parity violation and a light neutralino \cite{Dib_2023}. In addition, since we seek a charged final state, the result could also be reinterpreted as a search specifically for charged $B$-mesogenesis, as described in Ref.~\cite{Elahi:2021jia}. Charged $B$-mesogenesis scenarios are being actively probed at several collider-based and neutrino experiments. The result from the present work can provide a relevant constraint for these studies.


 The \babar~ detector is described in Refs.~\cite{BaBar:2001yhh,BaBar} and consists of several subsystems arranged in a cylindrical structure around the $e^{+}e^{-}$ interaction point. Charged-particle momenta are measured by a five-layer double-sided silicon vertex tracker and a 40-layer multi-wire drift chamber, both operating in the 1.5 T magnetic field of a superconducting solenoid. The particle identification (PID) for protons, kaons, and pions uses the specific energy loss measured in the tracking detectors and the measurement of the Cherenkov angle provided by the internally reflecting, ring-imaging Cherenkov detector. Photons are detected in the electromagnetic calorimeter (EMC). Muon identification is provided by the instrumented flux return. Protons are identified using ~\babar~ likelihood-based particle identification algorithms detailed in Ref.~\cite{lumi}. There is a negligible difference in the reconstruction efficiencies of protons and anti-protons. 
 
 The data sample used corresponds to an integrated luminosity of 398.5 fb$^{-1}$ \cite{lumi} collected at the PEP-II $e^{+}e^{-}$ storage ring at SLAC. A further 32.5 fb$^{-1}$ is used to optimize the analysis strategy and is excluded from the sample used to obtain the final result. At PEP-II, 9~GeV electrons collide with 3.1 GeV positrons at center-of-mass (CM) energies near 10.58 GeV ($\Upsilon(4S)$ resonance). The average cross section for $B^{+}B^{-}$ pair production of electron-positron annihilation is $ \sigma (e^{+}e^{-} \rightarrow B^{+} B^{-} ) \sim 550 $ pb; thus the data sample corresponds to $\sim 2 \times 10^{8}$ produced $B^{+}B^{-}$ pairs.


Monte Carlo (MC) generators are used to simulate background events that emanate from inclusive $e^{+}e^{-}\rightarrow B\bar{B}$ (EVTGEN \cite{EvtGen}) or  continuum $e^{+}e^{-}\rightarrow q\bar{q}   \: \:(q = udsc)$ processes (JETSET \cite{ref2,ref3}). Signal events are generated using EVTGEN. Samples were made for eight different $\psi_{\text{D}}$ mass hypotheses: 1.0, 1.5, 2.0, 2.5, 3.0, 3.5, 4.0 and 4.2 GeV/$c^{2}$. The propagation of particles through the detector is simulated using the GEANT4 toolkit \cite{Geant}. 


The reconstructed $B_{\text{tag}}$ must have a CM energy ($E^{*}_{B_{\text{tag}}}$) within $\pm$0.2 GeV of the beam energy, $E^{*}_{\text{beam}}$, in the CM frame. The energy-substituted mass is defined as $ m_{\text{ES}} c^{2}= \sqrt{E_{\text{beam}}^{\:* 2} - \vec{\text{p}}^{ \:* 2}_{B_{\text{tag}}} c^{2}}$, where $\vec{\text{p}}^{ \:*}_{B_{\text{tag}}}$ is the three-momentum of $B_{\text{tag}}$ in the CM frame. We require $m_{\text{ES}}$ of the  $B_{\text{tag}}$ to lie within the nominal $B^{+}$ mass range defined by $5.27-5.29$ GeV/c$^{2}$.  When multiple $B_{\text{tag}}$ candidates are found in one event, the one that has the lowest value of  $\Delta E = E^{*}_{\text{beam}} - E^{*}_{B_{\text{tag}}}$ is selected. 

 On the signal side, the presence of one and only one charged track is required, and it must be consistent with the proton hypothesis. To suppress the remaining inclusive background, we use a single multivariate classifier based on a boosted decision tree (BDT) algorithm which is trained on the combined background and signal MC samples. The BDT
includes the following kinematic variables from the $B_{\text{tag}}$: $\Delta E$ and ${m_{\text{ES}}}$; information about the hadronic decay channel and its purity \cite{BaBar:2014omp}; and the magnitude of the
thrust vector, defined as the sum of the magnitudes of the
momenta of all tracks and calorimeter clusters projected
onto the thrust axis \cite{BaBar:2014omp}. The following features from the $B_{\text{sig}}$ are also included: the total extra neutral energy on the signal side in the CM frame; the cosine of the polar angle of the missing momentum vector recoiling against the $B_{\text{tag}}$ meson and the signal candidate in the laboratory frame; the number of neutral particles and the number of $\pi^0$ candidates on the signal side, where a $\pi^0$ candidate is two photons with an invariant mass within $15$ MeV/c$^{2}$ of the nominal $\pi^0$ mass (134.9 MeV/c$^{2}$~\cite{PDG2020}). Additional features include the ratio of the second to zeroth Fox-Wolfram \cite{FW} moment for all tracks and neutral clusters (denoted as $R_{2}$), and the cosine of the thrust vector. These features are uncorrelated (in most cases $\ll 50 \%$) for both signal and background events. The features that provide the most discriminating power are $m_{ES}$, purity and decay information, $R_{2}$ and thrust vector magnitude. An additional criterion that no neutral pion candidates should present on the signal side is applied before the final analysis, at which point no extra neutral candidates remain.

\begin{figure}[t!]
    \centering
    \includegraphics[width=0.45\textwidth]{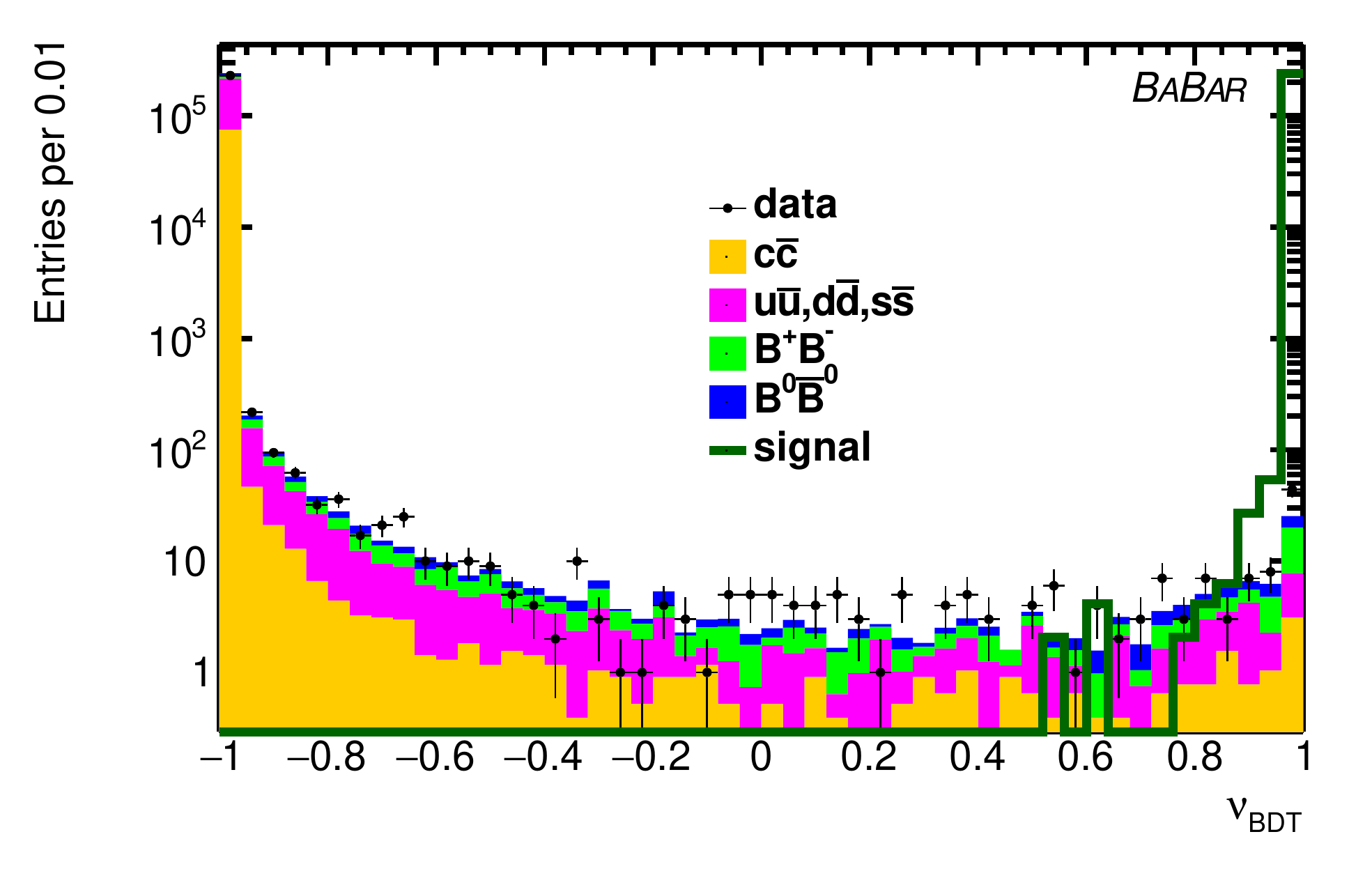}
    \caption{ BDT response for data and all backgrounds. The signal shown is an inclusive signal sample including all 8 simulated signal samples.}
    \label{fig:MVA5}
\end{figure}

Figure~\ref{fig:MVA5} shows the distribution of the BDT responses ($\nu_{\text{BDT}} $). Events are required to have $\nu_{\text{BDT}} > 0.95$, which retains $>$ 99$\%$ of all the simulated signals and 0.0028$\%$  of the simulated background.

\begin{figure}[t]
    \centering
    \includegraphics[width=0.45\textwidth]{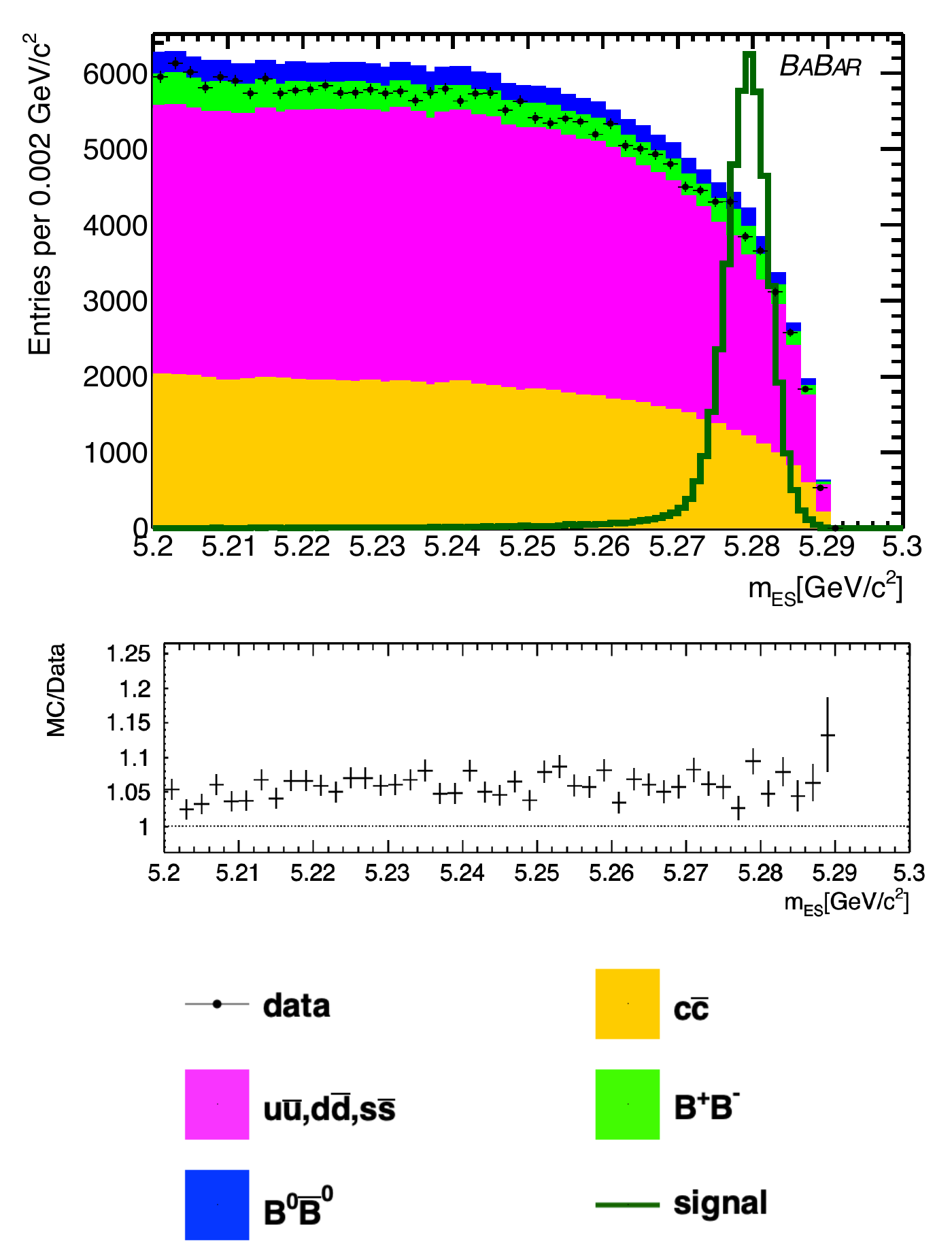}
    \caption{ The energy-substituted mass ($m_{\text{ES}}$) of the $B_{\text{tag}}$ candidate for MC background processes and data.  An example signal distribution is shown with arbitrary normalization (no correction applied).}
    \label{fig:ProtonMES}
\end{figure}

Figure~\ref{fig:ProtonMES} shows the distribution of $m_{\text{ES}}$ for inclusive MC background, signal, and data. The signal events peak around the nominal $B$ meson mass and background events are dominated by the continuum events. 

\begin{figure}[t]
    \centering
    \includegraphics[width=0.45\textwidth]{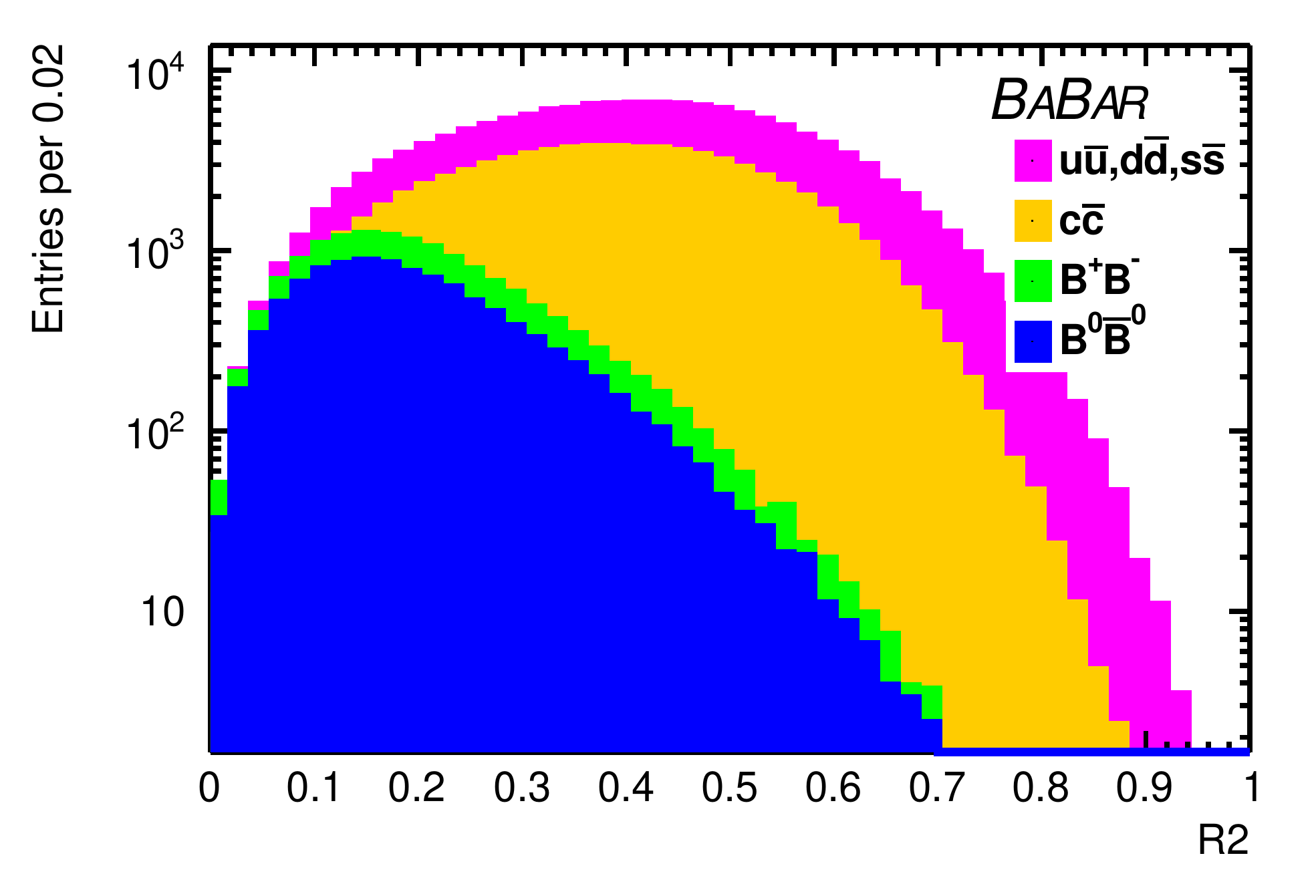}
    \caption{ Simulated distributions of the ratio of the second-to-zeroth Fox-Wolfram moment for all tracks (denoted as $R_{2}$).}
    \label{fig:allProtonR2}
\end{figure}

Known discrepancies in the simulation \cite{BaBar:2009pvj} of the $B\bar{B}$  and of $q\bar{q}$ events are corrected for, in a two-stage process, based on an analysis of the distribution of $R_{2}$ (Fig.~\ref{fig:allProtonR2}). First, a correction factor for the $q\bar{q}$ samples, $f_{q\bar{q}} = 1.05 \pm 0.03$, is extracted from  the $R_{2} > 0.7$ region.  Similarly, a correction factor, $f_{B\bar{B}} = 0.85 \pm 0.07$ for the $B\bar{B}$ samples is extracted from the  $R_{2} < 0.7$  region, assuming an equal contribution to the correction from both $B^{0}\bar{B}^{0}$ and $B^{+}\bar{B}^{-}$.  In both cases, the uncertainties are purely statistical. Under the assumption that $f_{B\bar{B}}$ is independent of the $B_{\text{sig}}$ decay mode, the signal efficiency is also re-scaled by $f_{B\bar{B}}$.


The signal efficiencies are extracted as the ratios of selected events to the total generated from the eight simulated signal samples. The signal efficiency varies from 0.00145 for $m_{\psi_{\text{D}}} = 1.0$ GeV/c$^{2}$ to 0.0006  for $m_{\psi_{\text{D}}} = 4.2$ GeV/c$^{2}$. The largest loss of efficiency comes from use of the standard ~\babar~reconstruction algorithm and the requirements of a proton track, with no accompanying charged particles on the signal side. The efficiencies extracted from the eight signal samples are fitted with a smooth seventh-order polynomial, with $\chi^{2}/ndf = 0.98$, to allow interpolation at any intermediate mass hypothesis.

\begin{figure}[t]
    \centering
    \includegraphics[width=0.45\textwidth,height=2.2in]{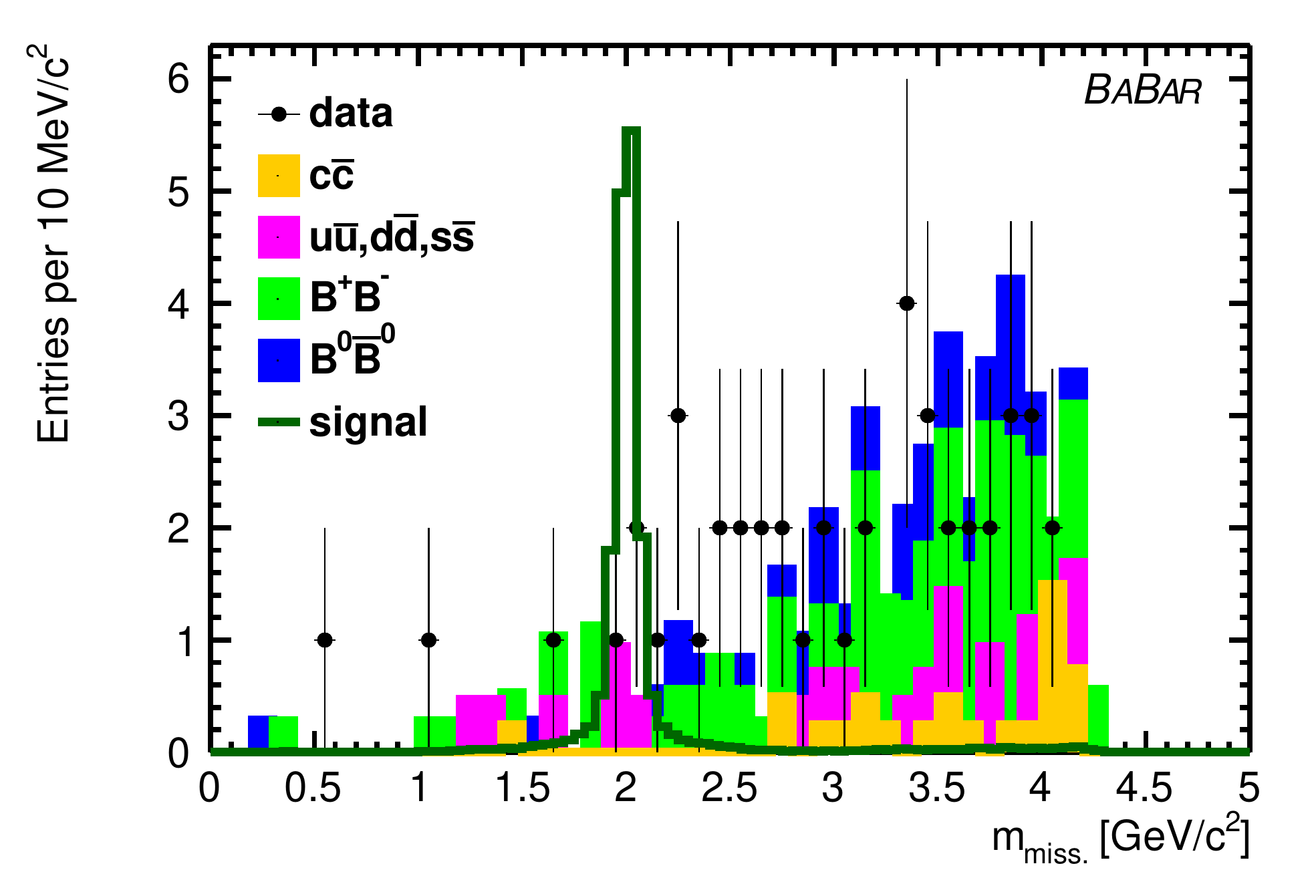}
    \caption{ Missing-mass distributions after all selections are applied for a simulated signal sample with $m_{\psi_{\text{D}}} = $ 2 GeV/c$^{2}$ (solid green line), inclusive SM background (stack histograms) and data (black dots). In total 46 events remain in the data and 48 remain in the inclusive SM MC sample.  }
    \label{fig:allProtonMissingMass}
\end{figure}

The missing-mass ($m_{\text{miss}}$), which in the case of a signal would be the $\psi_{\text{D}}$ mass, is calculated from
the four momenta of the signal $B_{\text{sig}}$ and proton:

\begin{equation}
    m_{\text{\text{miss}}} c^{2} = \sqrt{(E^{\:*}_{B_{\text{sig}}} - E^{\:*}_{\text{p}})^{2} - | \vec{\text{p}}^{\:*}_{B_{\text{sig}}} - \vec{\text{p}}^{\:*}_{\text{p}} |^{2} c^{2}}
\end{equation}
where ($\vec{\text{p}}^{\:*}_{B_{\text{sig}}},E^{\:*}_{B_{\text{sig}}}$) and ($\vec{\text{p}}^{\:*}_{\text{p}},E^{\:*}_{\text{p}}$) are the four-momenta of the signal $B_{\text{sig}}$ and proton, respectively, in the CM frame. Figure~\ref{fig:allProtonMissingMass} shows the missing-mass distribution for the data, background, and an example signal hypothesis after all selection criteria have been applied. For each signal mass, the missing-mass distribution is fitted with a double-sided Crystal Ball \cite{CB1,Skwarnicki:1986xj} function to extract the signal mass resolution.  The resolution is obtained from the fits to the signal MC and defined as $\sigma_{m}$ = FWHM/2.35; it varies from $\sim$110 MeV/c$^{2}$ at $m_{\text{\text{miss}}}$ = 1 GeV/c$^{2}$ to $\sim$11 MeV/c$^{2}$ at $m_{\text{miss}}$ = 4.2 GeV/c$^{2}$. The resolutions for all mass values in the search region are interpolated from the fit to the eight signal samples using an exponential function, the $\chi^{2}/ndf$ of the fit was 1.1.

A scan is performed across the missing-mass distribution with a step size equal to the signal mass resolution ($\sigma_{m}$) interpolated from fits to the signal MC samples. In total 127 mass hypotheses were considered in the range $1.0 <m_{\text{miss}} < 4.29$ GeV/c$^{2}$.

The largest systematic uncertainty comes from the data/MC correction (8.2$\%$)  and affects the signal efficiency. The uncertainty on the correction factor includes several contributions including imperfections in the modeling of reconstruction and particle identification. In addition, there are normalization uncertainties in the yield of $B^{+}B^{-}$ pairs which include: the uncertainty on the number of $\Upsilon(4S)$ mesons (0.6$\%$\cite{mcgregor2008b}); the uncertainty on the $\Upsilon(4S) \rightarrow B^{+}B^{-}$ branching fraction (1.2$\%$);
and, the uncertainty on the signal efficiency due to the PID algorithms incorrectly identifying a proton/anti-proton track (1$\%$). The total uncertainty on the signal efficiency is 8.4 $\%$.

In the absence of a signal, 90$\%$ confidence level (C.L.) upper limits on the branching fractions are derived using a profile likelihood method \cite{Rolke_2005}.  A Poisson counting approach is followed using only the data.  The number of signal and background events are assumed to follow Poisson distributions, and the efficiency is assumed Gaussian
with a standard deviation equal to the total systematic uncertainty. For a given $\psi_{\text{D}}$ mass hypothesis, the signal region is defined in the data as the region $m_{\psi_{\text{D}}} - 5\sigma_{m} < m_{\text{miss}} <  m_{\psi_{\text{D}}} + 5\sigma_{m}$, the side-bands ($[+5\sigma,+10\sigma]$ and $[-10\sigma,-5\sigma]$) on either side of this window are classified as the background region.

\begin{figure}[t]
    \centering
    \includegraphics[width=0.42\textwidth]{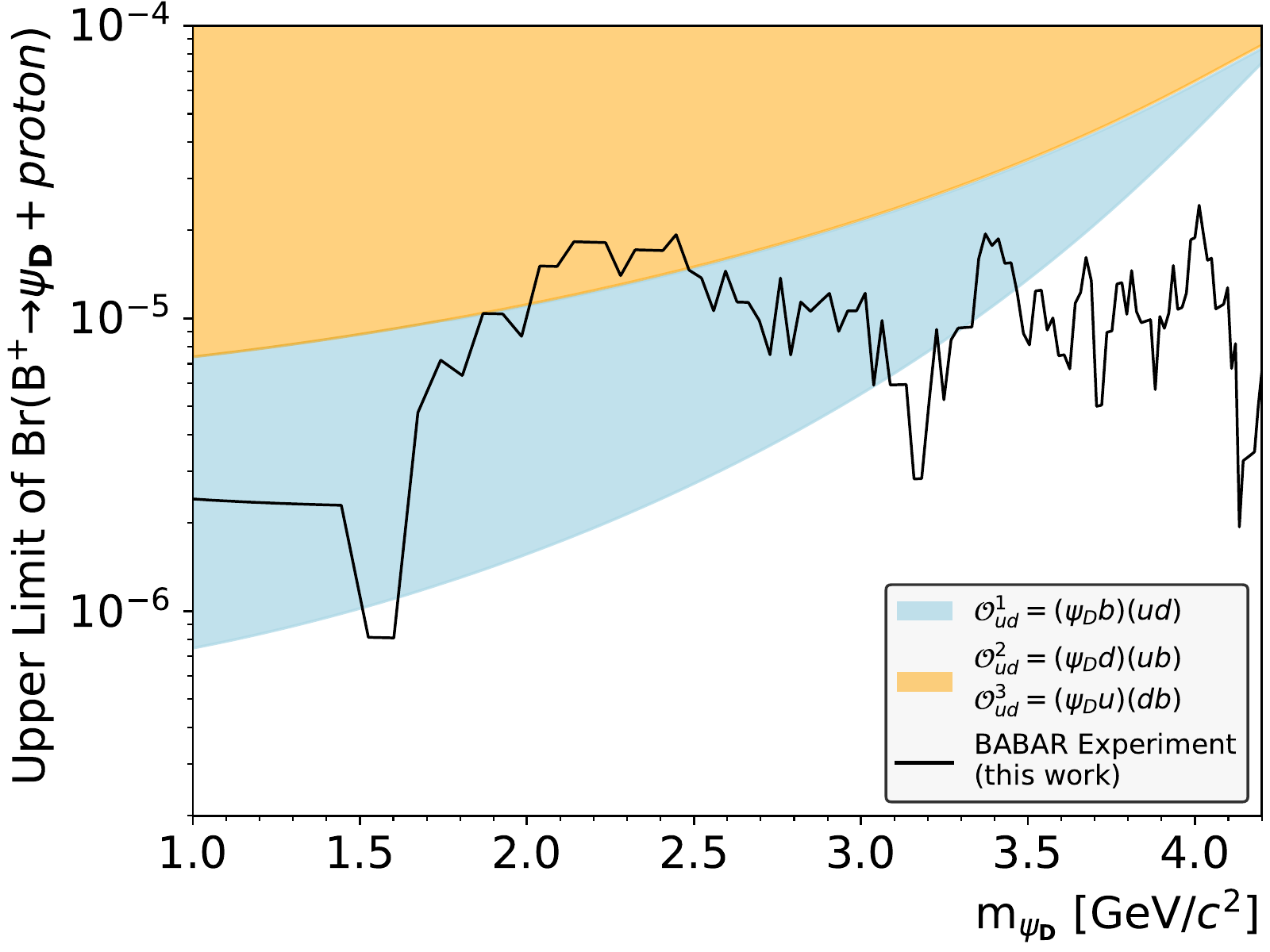}
    \caption{Derived 90 $\%$ C.L. upper limits on the branching fraction $B^{+} \rightarrow \psi_{\text{D}} + \text{p}$ and the charge conjugate for \babar~ data set corresponding to 398 fb$^{-1}$. The theory expectation for the three effective operators are from Ref. \cite{Alonso-Alvarez:2021qfd}.}
    \label{fig:protonlimit}
\end{figure}

Figure~\ref{fig:protonlimit} shows the resulting 90 $\%$ C.L. upper limit on the branching fraction. The largest local significance is 3.5 $\sigma$ at  3.3 GeV/c$^{2}$ which results in a 1 $\sigma$ global significance. Almost all the available parameter space for the $\mathcal{O}^{2,3}_{ud}$ operators is constrained with the \babar~ data set. However, operator $\mathcal{O}^{1}_{ud}$ remains mostly unconstrained between 1.9 -- 3.0 GeV/c$^{2}$ and below 1.5 GeV/c$^{2}$.

\begin{figure}[t]
    \centering
    \includegraphics[width=0.42\textwidth]{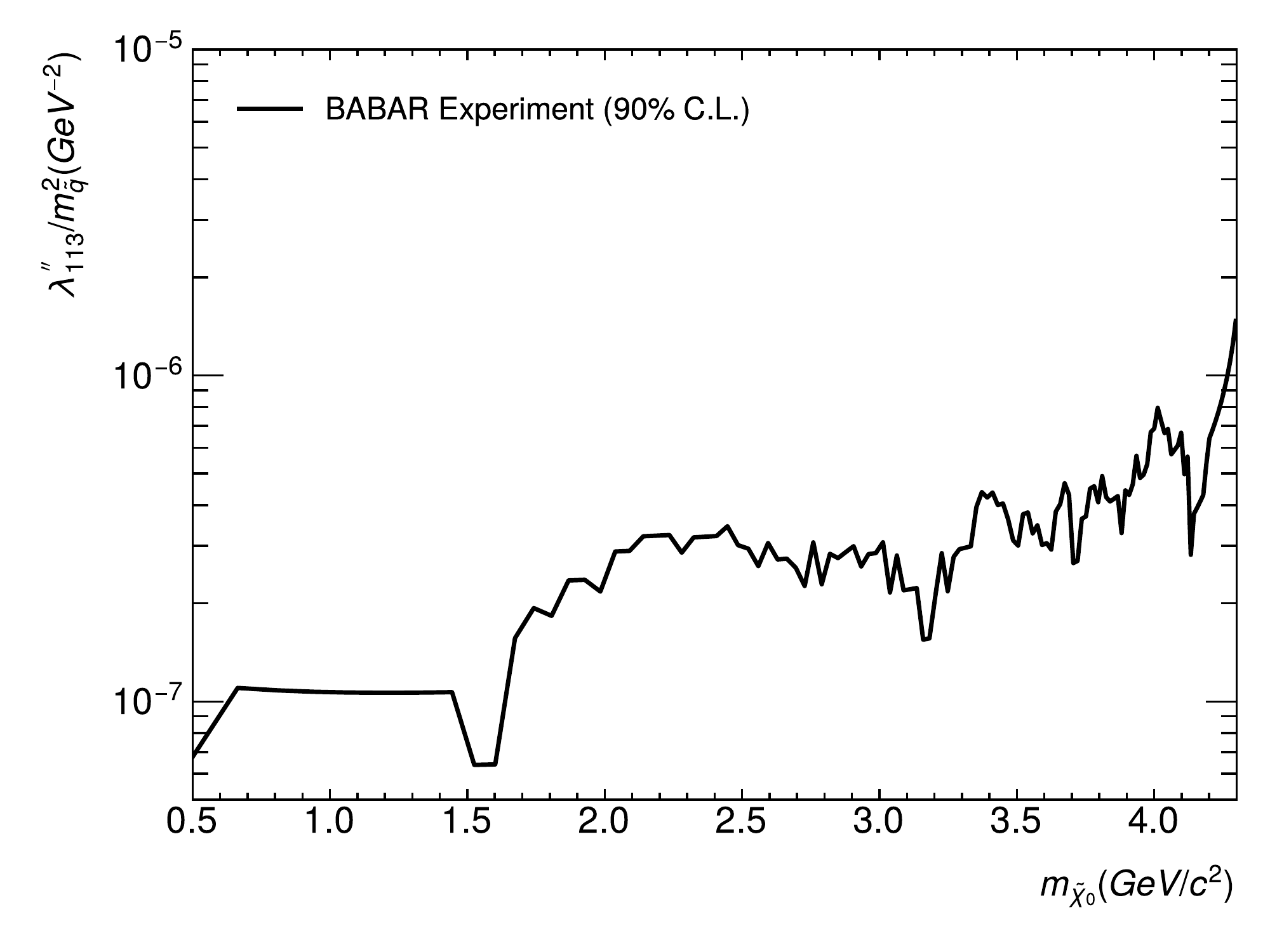}
    \caption{Derived 90 $\%$ C.L. upper limits  for \babar~ data set corresponding to 398 fb$^{-1}$ on the  RPV coupling $\lambda_{113}^{''}$ for the process $B^{+} \rightarrow \tilde{\chi}^{0} + p$ using the conversion factors presented in Fig. 2 of Ref. \cite{Dib_2023}. }
    \label{fig:SUSY}
\end{figure}

Our result can be reinterpreted to constrain other models with missing mass in the final state, including the RPV supersymmetry process $B^{+} \rightarrow \tilde{\chi}_{0} + p$, where $\tilde{\chi}_{0}$ is the lightest neutralino. In Fig.~\ref{fig:SUSY} the branching fraction upper limits obtained in the present analysis are converted to limits on the  RPV coupling $\lambda_{113}^{''}$ divided by the relevant squark mass squared as a function of the neutralino mass. These are unique limits, there are no previous results for this channel.

To summarize, a search for  $B^{+} \rightarrow \psi_{\text{D}} + \text{p}$  has been presented. This is the first attempt to directly search for this channel. No signal is observed, and 90$\%$ C.L upper limits from $10^{-7} $ -- $ 10^{-5}$ are set on the branching fraction. A large fraction of the $B$-mesogenesis parameter space is excluded by this measurement. Our result also constrains the branching fraction upper limits  on the  RPV coupling, $\lambda_{113}^{''}$, divided by the relevant squark mass squared as a function of the neutralino mass, at the level $10^{-7}$ -- $10^{-6}$ for 0.5$< m_{\tilde{\chi}^{0}} <$ 4.29 GeV/c$^{2}$. In addition, we note that the limits (outlined in the Supplementary Material) can also be reinterpreted to provide constraints on other models e.g. charged $B$-mesogenesis.

We are grateful for the extraordinary contributions of our PEP-II colleagues in achieving the excellent luminosity and machine conditions that have made this work possible. The success of this project also relies critically on the expertise and dedication of the computing organizations that support BABAR, including GridKa, UVic HEP-RC, CC-IN2P3, and CERN. The collaborating institutions wish to thank SLAC for its support and the kind hospitality extended to them. We also wish to acknowledge the important contributions of J.~Dorfan and our deceased colleagues E.~Gabathuler, W.~Innes, D.W.G.S.~Leith, A.~Onuchin, G.~Piredda, and R. F.~Schwitters.

\bibliography{paper}

\end{document}


\begin{table*}[t]
\centering
\caption{Results for the final data sample}
\begin{tabular}{c | c | c | c | c | c | c | c | c} 
 \hline 
bin & $m_{\psi_{DM}}$ & $\sigma_{\psi_{DM}}$ & $\mathcal{N}_{bkg}$ &  $\mathcal{N}_{sig}$ & $\epsilon_{MC}$ & $\delta \epsilon^{stat}_{MC}$ & $\delta \epsilon^{sys}_{MC}$ & $Br$ (90 $\%$ C.L ) \\ [0.5ex] 
\hline
\hline
1&0.5&0.1642&11&2&0.00111&1.029e-05&0.000243&9.993e-07\\
2&0.664&0.1451&9&2&0.00119&1.162e-05&0.000260&2.627e-06\\
3&0.809&0.1301&9&2&0.00124&1.239e-05&0.000271&2.515e-06\\
4&0.939&0.1181&9&2&0.00128&1.284e-05&0.000280&2.442e-06\\
5&1.058&0.1081&9&2&0.001311&1.312e-05&0.000285&2.393e-06\\
6&1.166&0.0997&9&2&0.00133&1.329e-05&0.000289&2.358e-06\\
7&1.265&0.0925&9&2&0.00134&1.340e-05&0.000292&2.333e-06\\
8&1.358&0.0863&9&2&0.00135&1.348e-05&0.000295&2.314e-06\\
9&1.444&0.0810&9&2&0.001360&1.353e-05&0.000297&2.299e-06\\
10&1.525&0.0762&11&1&0.00137&1.356e-05&0.000298&8.135e-07\\
11&1.601&0.0720&11&1&0.00137&1.359e-05&0.000300&8.098e-07\\
12&1.674&0.0683&9&3&0.00138&1.362e-05&0.000301&4.768e-06\\
13&1.741&0.0649&9&4&0.00139&1.365e-05&0.000302&7.201e-06\\
14&1.807&0.0618&10&4&0.00139&1.367e-05&0.000303&6.377e-06\\
15&1.869&0.0591&8&5&0.00140&1.370e-05&0.000304&1.039e-05\\
16&1.928&0.0565&8&5&0.00140&1.372e-05&0.000305&1.036e-05\\
17&1.984&0.0542&10&5&0.00141&1.3746e-05&0.000306&8.681e-06\\
18&2.0383&0.0520&8&7&0.00141&1.377e-05&0.000307&1.510e-05\\
19&2.0904&0.0501&8&7&0.00142&1.380e-05&0.000309&1.505e-05\\
20&2.1405&0.0482&7&8&0.00142&1.382e-05&0.000310&1.828e-05\\
21&2.189&0.0466&7&8&0.00142&1.385-05&0.000311&1.822e-05\\
22&2.235&0.04501&7&8&0.00143&1.387e-05&0.000312&1.816e-05\\
23&2.280&0.04354&9&7&0.001434&1.390e-05&0.000313&1.400e-05\\
24&2.324&0.04217&8&8&0.001439&1.392e-05&0.000314&1.716e-05\\
25&2.366&0.04090&8&8&0.00144&1.395e-05&0.0003151&1.710e-05\\
26&2.407&0.03967&8&8&0.00145&1.397e-05&0.000316&1.705e-05\\
27&2.446&0.03853&8&9&0.00145&1.399e-05&0.000317&1.934e-05\\
28&2.485&0.03745&8&7&0.00146&1.401e-05&0.000317&1.463e-05\\
29&2.523&0.03643&9&7&0.00146&1.403e-05&0.000318&1.375e-05\\
30&2.559&0.03547&10&6&0.00146&1.405e-05&0.000319&1.061e-05\\
31&2.595&0.03456&8&7&0.00147&1.407e-05&0.000320&1.453e-05\\
32&2.629&0.03370&9&6&0.00147&1.408e-05&0.000320&1.137e-05\\
33&2.663&0.03288&9&6&0.00147&1.410e-05&0.000321&1.135e-05\\
34&2.696&0.03209&8&5&0.00148&1.411e-05&0.000322&9.836e-06\\
35&2.728&0.03134&8&4&0.00148&1.412e-05&0.000322&7.513e-06\\
36&2.759&0.03064&6&6&0.00148&1.413e-05&0.000323&1.375e-05\\
37&2.790&0.02996&8&4&0.00148&1.414e-05&0.000323&7.495e-06\\
38&2.820&0.02931&6&5&0.00148&1.415e-05&0.000323&1.139e-05\\
39&2.849&0.02869&7&5&0.00148&1.416e-05&0.000323&1.058e-05\\
40&2.878&0.02810&6&5&0.00148&1.416e-05&0.000324&1.137e-05\\
41&2.906&0.02753&5&5&0.00148&1.416e-05&0.000324&1.218e-05\\
42&2.933&0.02699&6&4&0.00148&1.417e-05&0.000324&9.021e-06\\
43&2.960&0.02647&4&4&0.00148&1.416e-05&0.000324&1.061e-05\\
44&2.987&0.02596&4&4&0.00148&1.416e-05&0.000323&1.061e-05\\
45&3.013&0.02547&5&5&0.00148&1.416e-05&0.000323&1.220e-05\\
46&3.038 &0.0251&7&3&0.00148&1.415e-05&0.000323&5.910e-06\\
47&3.063&0.02456&5&4&0.00148&1.415e-05&0.000322&9.841e-06\\
48&3.087&0.02413&7&3&0.00148&1.414e-05&0.000322&5.925e-06\\
49&3.112&0.02371&7&3&0.00147&1.413e-05&0.000322&5.935e-06\\
50&3.136&0.02331&7&3&0.00147&1.412e-05&0.000321&5.945e-06\\
51&3.159&0.02291&8&2&0.00147&1.411e-05&0.000320&2.828e-06\\
52&3.182&0.02254&8&2&0.00147&1.409e-05&0.000320&2.835e-06\\
53&3.204&0.02218&8&3&0.00146&1.408e-05&0.000319&5.239e-06\\
54&3.227&0.02183&6&4&0.00146&1.406e-05&0.000318&9.184e-06\\
55&3.248&0.0215&8&3&0.00145&1.404e-05&0.000317&5.269e-06\\
56&3.270&0.02116&7&4&0.00145&1.402e-05&0.000316&8.446e-06\\
57&3.291&0.02084&6&4&0.00144&1.400e-05&0.000315&9.273e-06\\
58&3.312&0.02053&6&4&0.00144&1.398e-05&0.000314&9.307e-06\\
59&3.332&0.02023&6&4&0.00143&1.395e-05&0.000312&9.344e-06\\
60&3.353&0.01993&4&6&0.00143&1.392e-05&0.0003101&1.601e-05\\
61&3.373&0.01965&3&7&0.00142&1.390e-05&0.000310&1.946e-05\\
62&3.392&0.01938&5&7&0.00141&1.387e-05&0.000308&1.773e-05\\
63&3.412&0.01911&4&7&0.001404&1.384e-05&0.000307&1.873e-05\\
\end{tabular}
\label{table:run3}
\end{table*}

\begin{table*}[t!]
\centering
\begin{tabular}{c | c | c | c | c | c | c | c | c} 
bin & $m_{\psi_{DM}}$ & $\sigma_{\psi_{DM}}$ & $\mathcal{N}_{bkg}$ &  $\mathcal{N}_{sig}$ & $\epsilon_{MC}$ & $\delta \epsilon^{stat}_{MC}$ & $\delta \epsilon^{sys}_{MC}$ & $Br$ (90 $\%$ C.L ) \\ [0.5ex] 
\hline
\hline
64&3.431&0.01885&5&6&0.00140&1.380e-05&0.000305&1.541e-05\\
65&3.450&0.01860&5&6&0.00139&1.377e-05&0.000304&1.550e-05\\
66&3.468&0.01835&6&5&0.00139&1.374e-05&0.000302&1.219e-05\\
67&3.487&0.01811&7&4&0.00138&1.370e-05&0.000300&8.890e-06\\
68&3.505&0.01788&8&4&0.001369&1.366e-05&0.000298&8.110e-06\\
69&3.523&0.01766&6&5&0.00136&1.362e-05&0.000297&1.241e-05\\
70&3.540&0.01744&6&5&0.00135&1.358e-05&0.000295&1.249e-05\\
71&3.558&0.01722&7&4&0.00134&1.354e-05&0.000293&9.120e-06\\
72&3.575&0.01701&6&4&0.00133&1.349e-05&0.000299&1.005e-05\\
73&3.592&0.01681&6&3&0.00132&1.345e-05&0.000288&7.455e-06\\
74&3.609&0.01661&6&3&0.00131&1.341e-05&0.000286&7.511e-06\\
75&3.625&0.01641&7&3&0.00130&1.335e-05&0.000284&6.720e-06\\
76&3.642&0.01623&5&4&0.00129&1.330e-05&0.000282&1.127e-05\\
77&3.658&0.01604&4&4&0.00128&1.325e-05&0.000279&1.228e-05\\
78&3.674&0.01585&3&5&0.00127&1.320e-05&0.000277&1.617e-05\\
79&3.690&0.01568&3&4&0.00126&1.315e-05&0.000275&1.345e-05\\
80&3.706&0.01551&6&2&0.00125&1.309e-05&0.000272&5.009e-06\\
81&3.721&0.01534&6&2&0.00124&1.303e-05&0.000270&5.056e-06\\
82&3.736&0.01517&5&3&0.00123&1.297e-05&0.000267&8.969e-06\\
83&3.752&0.01501&5&3&0.00121&1.291e-05&0.000264&9.058e-06\\
84&3.767&0.01485&4&4&0.00120&1.285e-05&0.000262&1.311e-05\\
85&3.781&0.01470&4&4&0.00119&1.279e-05&0.000259&1.325e-05\\
86&3.796&0.01455&4&3&0.00118&1.273e-05&0.000256&1.032e-05\\
87&3.811&0.01440&3&4&0.00116&1.266e-05&0.000253&1.458e-05\\
88&3.825&0.01426&4&3&0.00115&1.259e-05&0.000251&1.055e-05\\
89&3.839&0.01411&5&3&0.00114&1.252e-05&0.000248&9.670e-06\\
90&3.853&0.01397&5&3&0.00112&1.245e-05&0.000245&9.786e-06\\
91&3.867&0.01384&5&3&0.00111&1.238e-05&0.000242&9.906e-06\\
92&3.881&0.01370&6&2&0.00110&1.231e-05&0.000239&5.710e-06\\
93&3.895&0.01357&5&3&0.00108&1.223e-05&0.000236&1.016e-05\\
94&3.908&0.01345&6&3&0.00107&1.216e-05&0.000233&9.244e-06\\
95&3.922&0.01332&5&3&0.00105&1.208e-05&0.000230&1.044e-05\\
96&3.935&0.01320&4&4&0.00104&1.200e-05&0.000226&1.517e-05\\
97&3.948&0.01308&5&3&0.00102&1.192e-05&0.000223&1.074e-05\\
98&3.962&0.01296&5&3&0.00102&1.184e-05&0.000220&1.089e-05\\
99&3.974&0.01284&4&3&0.00099&1.176e-05&0.0002171&1.221e-05\\
100&3.987&0.01273&2&4&0.00098&1.167e-05&0.000213&1.857e-05\\
101&4.000&0.01261&2&4&0.00096&1.159e-05&0.000210&1.886e-05\\
102&4.013&0.01250&1&5&0.00095&1.150e-05&0.000206&2.436e-05\\
103&4.025&0.01240&2&4&0.00093&1.141e-05&0.000203&1.950e-05\\
104&4.038&0.01229&2&3&0.00092&1.132e-05&0.000200&1.579e-05\\
105&4.050&0.01219&2&3&0.00090&1.123e-05&0.000196&1.607e-05\\
106&4.062&0.01208&3&2&0.00088&1.113e-05&0.000193&1.077e-05\\
107&4.074&0.0120&3&2&0.00087&1.104e-05&0.000189&1.0967e-05\\
108&4.086&0.01188&3&2&0.00085&1.094e-05&0.000186&1.118e-05\\
109&4.100&0.01179&2&2&0.00083&1.085e-05&0.000182&1.274-05\\
110&4.111&0.01169&3&1&0.00082&1.074e-05&0.000178&6.746e-06\\
111&4.125&0.01160&2&1&0.00080&1.064e-05&0.000175&8.201e-06\\
112&4.133&0.01151&3&0&0.00078&1.054e-05&0.000171&1.940e-06\\
113&4.145&0.01141&2&0&0.00077&1.044e-05&0.000167&3.265e-06\\
114&4.156&0.01132&2&0&0.00075&1.033e-05&0.000164&3.339e-06\\
115&4.167&0.01124&2&0&0.00073&1.0227e-05&0.000160&3.418e-06\\
116&4.179&0.01115&2&0&0.00072&1.0119e-05&0.000157&3.500e-06\\
117&4.190&0.01107&1&0&0.00070&1.001e-05&0.000152&5.0248e-06\\
118&4.201&0.01098&0&0&0.00068&9.900e-06&0.000148&6.658e-06\\
119&4.212&0.01090&0&0&0.00066&9.786e-06&0.000145&6.835e-06\\
120&4.223&0.01082&0&0&0.00065 &9.670e-06&0.000141&7.023e-06\\
121&4.234&0.01074&0&0&0.00063&9.560e-06&0.000137&7.222e-06\\
122&4.244&0.01066&0&0&0.00061&9.440e-06&0.000133&7.434e-06\\
123&4.255&0.01058&0&0&0.00059&9.327e-06&0.000129&7.660e-06\\
124&4.265&0.01051&0&0&0.00057&9.202e-06&0.000125&7.901e-06\\
125&4.276&0.01043&0&0&0.00056&9.081e-06&0.000121&8.160e-06\\
126&4.286&0.01036&0&0&0.00054&8.958e-06&0.000117&8.437e-06\\
127&4.297&0.01028&0&0&0.00052&8.833e-06&0.000113&8.735e-06\\
\hline
\end{tabular}
\label{table:allrun3}
\end{table*}